\documentclass[conference]{IEEEtran}
\IEEEoverridecommandlockouts

\usepackage{amsmath,amssymb,amsfonts}
\usepackage[ruled]{algorithm2e}
\usepackage{algorithmic}
\usepackage{graphicx}
\usepackage{textcomp}
\usepackage{xcolor}
\usepackage{hyperref}
\usepackage[sort, numbers]{natbib}
\usepackage{caption}
\usepackage{subcaption}

\usepackage{color}
\usepackage{listings}
\definecolor{GrayCodeBlock}{RGB}{241,241,241}
\definecolor{BlackText}{RGB}{110,107,94}
\definecolor{RedTypename}{RGB}{182,86,17}
\definecolor{GreenString}{RGB}{96,172,57}
\definecolor{PurpleKeyword}{RGB}{184,84,212}
\definecolor{GrayComment}{RGB}{170,170,170}
\definecolor{GoldDocumentation}{RGB}{180,165,45}
\lstdefinelanguage{rust}
{
    columns=fullflexible,
    keepspaces=true,
    frame=single,
    framesep=0pt,
    framerule=0pt,
    framexleftmargin=4pt,
    framexrightmargin=4pt,
    framextopmargin=5pt,
    framexbottommargin=3pt,
    xleftmargin=4pt,
    xrightmargin=4pt,
    basicstyle=\ttfamily\color{BlackText},
    keywords={
        true,false,
        unsafe,async,await,move,
        use,pub,crate,super,self,mod,
        struct,enum,fn,const,static,let,mut,ref,type,impl,dyn,trait,where,as,
        break,continue,if,else,while,for,loop,match,return,yield,in
    },
    keywordstyle=\color{PurpleKeyword},
    ndkeywords={
        bool,u8,u16,u32,u64,u128,i8,i16,i32,i64,i128,char,str,
        Self,Option,Some,None,Result,Ok,Err,String,Box,Vec,Rc,Arc,Cell,RefCell,HashMap,BTreeMap,
        macro_rules
    },
    ndkeywordstyle=\color{RedTypename},
    comment=[l][\color{GrayComment}\slshape]{//},
    morecomment=[s][\color{GrayComment}\slshape]{/*}{*/},
    morecomment=[l][\color{GoldDocumentation}\slshape]{///},
    morecomment=[s][\color{GoldDocumentation}\slshape]{/*!}{*/},
    morecomment=[l][\color{GoldDocumentation}\slshape]{//!},
    morecomment=[s][\color{RedTypename}]{\#![}{]},
    morecomment=[s][\color{RedTypename}]{\#[}{]},
    stringstyle=\color{GreenString},
    string=[b]"
}

\def\BibTeX{{\rm B\kern-.05em{\sc i\kern-.025em b}\kern-.08em
    T\kern-.1667em\lower.7ex\hbox{E}\kern-.125emX}}

\newtheorem{theorem}{Theorem}
\newtheorem{proposition}[theorem]{Proposition}

\begin{document}

\title{Petri Nets for Concurrent Programming}

\author{\IEEEauthorblockN{Marshall Rawson}
\IEEEauthorblockA{
\textit{University of Florida}\\
Gainesville, FL \\
marshallrawson@ufl.edu}
\and
\IEEEauthorblockN{Michael G. Rawson}
\IEEEauthorblockA{
\textit{Pacific Northwest National Lab}\\
Seattle, WA \\
michael.rawson@pnnl.gov}
}

\maketitle

\begin{abstract}
Concurrent programming is used in all large and complex computer systems. However, concurrency errors and system failures (ex: crashes and deadlocks) are common. We find that Petri nets can be used to model concurrent systems and find and remove errors ahead of time.
We introduce a novel generalization of Petri nets with nondeterministic transition nodes to match real systems. These allow for a compact way to construct, optimize, and prove computer programs at the concurrency level. 
Petri net programs can also be optimized by automatically solving for maximal concurrency, where the maximum number of valid threads is determined by the structure of the Petri net prior to execution. 
We discuss an algorithm to compute the state graph of a given Petri net start state pair.
We introduce our open source software 
\href{https://github.com/MarshallRawson/nt-petri-net}{framework}\footnote{https://github.com/MarshallRawson/nt-petri-net}
which implements this theory as a general purpose concurrency focused middle-ware.

\end{abstract}


\section{Introduction}

\subsection{What is a Petri Net}

Petri nets are named after the inventor Carl Petri in 1939 for the purposes of describing chemical processes \cite{petri}.
A Petri net is a bipartite directed graph with tokens assigned to nodes. A Petri net operates by letting tokens move following a set of rules.
The two partitions of nodes in a Petri net are called place nodes and transition nodes. As a bipartite graph, place nodes link to transition nodes and vice versa but place nodes do not connect to each other and neither do transition nodes. 
Only place nodes can hold tokens.
A place node can hold any number of tokens.
Each directed edge has a positive integer weight.
A transition node becomes enabled if all of the place nodes that point to it have more tokens than the weight on their corresponding edges.
A transition node will fire once it is enabled. 
Upon firing, the transition node's incoming edges removes that many tokens from the corresponding place and the outgoing edges add that many tokens to the corresponding place. 

\begin{figure}[h]
    \centering
    
     \begin{subfigure}[b]{0.4\textwidth}
         \centering
         
        \includegraphics[scale=0.7]{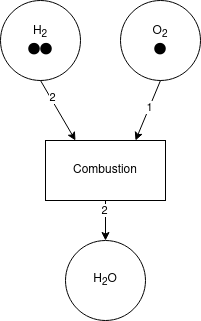}
        \caption{Initial state before ``firing".}
        \label{fig:h2o_0}
     \end{subfigure}

     \begin{subfigure}[b]{0.4\textwidth}
         \centering

        \includegraphics[scale=0.7]{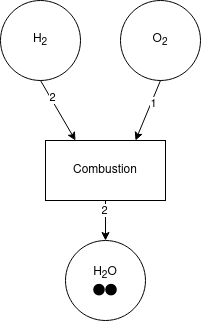}
        \caption{Ending state after ``firing".}
        \label{fig:h2o_1}
     \end{subfigure}
    \caption{A Petri net modeling $2H_2 + O_2 \rightarrow 2H_{2}O$. The circles are place nodes and the square is a transition node.}
    \label{fig:h2o}
\end{figure}

Note that the number of tokens is not necessarily conserved. For example, we model a chemical reaction with a Petri net in Figure \ref{fig:h2o}. 
Here, the firing corresponds to the chemical reaction and the tokens keep track of the atoms and compounds. 

\begin{figure}
    \centering
    \includegraphics[scale=0.7]{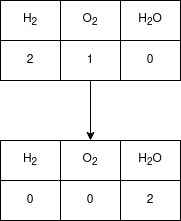}
    \caption{State graph of Petri net from Figure \ref{fig:h2o}.}
    \label{fig:h2o_state_graph}
\end{figure}

\begin{figure}
    \centering
    \includegraphics[scale=0.7]{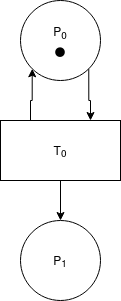}
    \caption{Unbounded Petri net start state pair.}
    \label{fig:unbounded_petri_net}
\end{figure}

\begin{figure}
    \centering
    \includegraphics[scale=0.7]{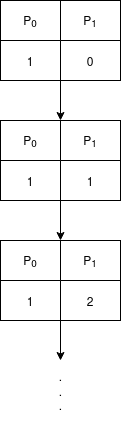}
    \caption{State graph of Figure \ref{fig:unbounded_petri_net}.}
    \label{fig:infinite_state_graph}
\end{figure}

The state of a given Petri net is described by the allocation of tokens to its place nodes.
Given a Petri net and a state, each possible next state may be found by applying enabled firing operations. 
From a starting state, a directed state graph can be constructed where each node is a state and each directed edge between states represents an enabled transition firing.
We will call a Petri net and an initial state, which is an allocation of tokens to place nodes, a Petri net start state pair.
A bounded Petri net start state pair has a finite number of nodes on its state graph and an unbounded Petri net start state pair has an infinite number of nodes on its state graph.
We show the state graph of the above Petri net start state pair in Figure \ref{fig:h2o_state_graph}.
We show an example of an unbounded Petri net start state pair in Figure \ref{fig:unbounded_petri_net} and its corresponding state graph in \ref{fig:infinite_state_graph}.

\subsection{Colored Petri Nets}

Colored Petri nets are a generalization of Petri nets described by C. R. Zeros in 1977 \cite{zeros}.
In a colored Petri net, each token has an associated color. Also, transition nodes are only enabled by sufficient tokens of certain colors as specified by the transition node. In the case that all tokens have the same color and transition nodes require the universal color, this is just a standard Petri net.

\begin{figure}
    \centering
    \includegraphics[scale=0.7]{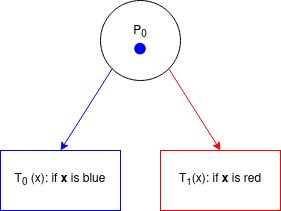}
    \caption{Colored Petri net with start tokens.}
    \label{fig:colored_petri_net}
\end{figure}

\begin{figure}
    \centering
    \includegraphics[scale=0.7]{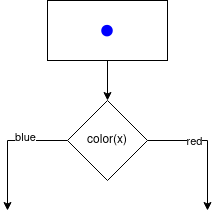}
    \caption{How Figure \ref{fig:colored_petri_net} maps onto a flow chart.}
    \label{fig:colored_pnet_flow_chart}
\end{figure}

Consider the example of a colored Petri net start state pair in Figure \ref{fig:colored_petri_net}. In this net, transition node $P_0$ has only a blue token. Transition node $T_0$ requires a blue token and transition node $T_1$ requires a red token. So, in the given start state, only the transition $T_0$ can fire.
Colored Petri nets which allow an infinite number of possible colors are Turing complete \cite{peterson}. However, even colored Petri nets with a finite number of colors can encode logic concisely.
Consider the colored Petri net start state pair Figure \ref{fig:colored_petri_net} where the only possible colors are red and blue. This colored Petri net can be expressed by the flow chart in Figure \ref{fig:colored_pnet_flow_chart}

\subsection{Contributions}

In Section \ref{sec:ntpetri},
we propose a generalization of Petri nets called Nondeterministic Transitioning Petri nets (NT-Petri nets) which allow for transitions to nondeterministically fire according to arbitrary firing conditions.
A NT-Petri net has a solvable state graph but allows the system designer to build systems with fewer elements on the graph. This is achieved by allowing each transition to have multiple state deltas determined by either hidden mechanisms or the particular distribution of tokens at the input places. We find that this greatly enhances designability of systems described with NT-Petri nets compared to standard Petri nets.

In Sections \ref{sec:concurrprog} and \ref{sec:concurrexec},
we introduce concurrent programs and discuss when they are useful or necessary for various system designs. Then we describe how the Petri nets model concurrent programs and contribute to the design process. 
We discuss how to concurrently execute a Petri net program. We also discuss the balance between concurrent and sequential execution of a given Petri nets from fully serialized to maximally concurrent.

In Section \ref{sec:software},
we introduce our open source software framework, available on \href{https://github.com/MarshallRawson/nt-petri-net}{Github.com}, that enables system designers to construct, execute, analyze, and optimize concurrent programs as NT-Petri nets. The framework is written in the well-known language Rust \cite{klabnik2019rust} for efficiency and compile-time memory safety. Then we use our open source software framework to build an example NT-Petri net featuring several firing conditions with nondeterministic outcomes due to hidden internal states or race condition nondeterminism.

In Section \ref{sec:cameraex},
we discuss a realistic use case for a NT-Petri net program. The use case is a processing pipeline with a feedback mechanism. The processing pipeline uses a pan tilt zoom camera and microphone to isolate video and audio of participants in a meeting. The feedback component of the processing pipeline is the pan tilt zoom commands that are fed back into the camera motor only after a later stage in the pipeline which determines which person to focus the camera on.
The NT-Petri net description of the system allows us to prove that the pipeline will never deadlock given some assumptions about the computations taking place. The assumptions are that the computations in the transitions always eventually finish and never cause the program to crash. We model the tunable parameters as NT-Petri net start tokens which to change the timing characteristics of the program.

\section{Related Previous Work}

A Petri net program is a computer program correctly modeled by a Petri net. A computer program is correctly modeled as a Petri net when the state of all concurrent elements are modeled with tokens and all state transitions, which map to computation happening on the tokens, are guaranteed to complete.
In 1998, a visual based Petri net programming language was built by Usher, M. and Jackson, D. \cite{usher}. It was successful in that it allowed the user to build and execute a computer program as a Petri net. However, it had some short comings in that the Petri net programs were not analyzable, completely visual based, did not have color, and used a slightly different definition of Petri net to allow for native composition of multiple nets.

While concurrent programming frameworks have been implemented and used for many years, there has always been a lagging behind of provability aspects of the final multiprograms \cite{siegel}. This is because the frameworks built and used thus far have not adhered to provable theoretic specifications. However, Petri nets are a provable theoretic standard which a middle-ware or framework can implement.

A framework similar to Petri nets has been designed with high throughput signal processing in mind as an implementation of the Kahn Process Network model \cite{allen}. While this framework is provably deterministic, it attempts to detect and resolve deadlocks at run time and needs to know that certain operations which are to be performed on the signal, are communicative in order to function, and is thus less general than Petri net frameworks.

Perennial is a system designed to machine verify concurrent programs which can crash, but only for a specific subset of the Go language (Goose) \cite{chajed}. In Perennial, the system designer provides source code, a specification, and a proof that the source code meets that specification where a computer program verifies the proof. This is a powerful system because it directly verifies concurrent source code which is allowed to crash, but it is currently language specific and has to implement versioned memory.

A known practical issue with Petri nets is state explosion. State explosion refers to the exponential growth of the state space of a Petri net with respect to the number of nodes and tokens in the Petri net. However, there has been work done in the past to counteract this by analyzing sections of Petri nets to independently solve for local state graphs, which are combined together later \cite{christensen}.

Another known shortcoming of Petri nets is that they do not model the duration of state transitions. Previous work has been done to slightly generalize Petri nets via timed Petri nets \cite{sifakis}. In a timed Petri net, transition firings are atomic, but tokens take time to become available once in their new places. This is plausibly useful when implementing Petri-net-based frameworks to capture the actual time taken to process tokens being computed in firing transitions.

There have been similar graph based computational languages, the most notable of them being the Data Flow Procedure Language \cite{Dennis1974First}. In this language, Dennis outlines a language where every fundamental arithmetic operation and conditional branch is represented by nodes in a graph connected by directed edges and are parallelizable. While this language is complete and can result in more compact and more parallelizable complete descriptions of a computation, it has four major draw backs: synchronization, complexity, performance, and integration.

The synchronization aspect of static dataflows as Dennis defined them have an implicit limit of one token per arc and all tokens have the same size. Culler expands on this by considering each token a pointer to a larger value in storage \cite{Culler}. However, this brings up the issue of needing to either define synchronization mechanisms in a dataflow to ensure the data in storage is not written to or read in the wrong order, or make all values pointed to by tokens copy-on-write. Culler also points out that there is no way in hardware to guarantee that there will always be one token per arc with the proposed hardware or software architecture, and proposes a change to add an acknowledgement arc to each data flow arc, which would increase the number of tokens exchanged by 1.5 to 2 times in a given dataflow.

The language is highly complex and has two types of fundamental tokens which are signals and data. The language has many types of nodes to handle all possible fundamental interactions between these types of tokens. This is in stark contrast to Petri nets and NT-Petri nets, which we propose to simply describe the system of computation to be executed in parallel. A high level language named Id was compiled into dataflow graphs and was executed on some novel CPUs \cite{Arvind1990}.

Performance will suffer if serialization is never exploited. If this language were to be executed in parallel on a multicore von Neumann computer, it would likely be much slower than the same computation evaluated by a completely serial program due to the incurred overhead of scheduling every operation through the operating system. A non-traditional CPU architecture has been proposed \cite{Dennis1974Preliminary} which would work specifically on data flow graphs, but it requires two additional large blocks of circuitry: a distribution network and an arbitration network, which would do the work of scheduling these parallelizable operations in hardware. There have been modifications to von Neumann CPUs \cite{Nikhil} to perform the execution and scheduling of data flow graphs on hardware. However, CPU design is a complex field, and it remains to be seen if these really have significant performance benefits over a multicore von Neumann machine which runs several serial programs that trade data and synchronize only on occasion. A completely parallelizable language is also not very integrable with the millions of existing computer systems with CPUs that are optimized for a small number of rapidly executed, mostly serial programs. Moreover, there do not exist large, well tested libraries written in this language for a potential system designer to leverage in building the computer program of interest.

In \cite{best1987sequential}, Petri nets are used to compare two different concurrency models: partial ordering and interleaved atomic instructions. However, in our NT-Petri net implementation, a combination is used where multiple work clusters are run in separate threads of execution. The work clusters execute regions of NT-Petri nets in a partial order. The work clusters are executed in separate threads, which may have interleaved execution according to the operating system scheduler.

\section{Nondeterministic Transitioning Petri Nets}\label{sec:ntpetri}

Deterministic programs have been the focus of previous work. However, now systems are sufficiently complex that they are designed to be nondeterministic. Nondeterministic programs have also been enabled by advances in programming languages that make it fast and easy to develop. For example, merely loading a webpage is now done nondeterministically, which greatly improves performance. However, when nondeterministic programs fail, they fail spectacularly and are much harder to fix and debug. We propose modelling nondeterministic programs with Petri nets. This novel idea allows the modeler to control and partition nondeterminism as we explain below.

Specifically, we will work with a more general version of colored Petri nets where transition nodes take tokens from a subset of its incoming place nodes and produce nondeterministic colored tokens to a nondeterministic subset of its outgoing place nodes. We will abbreviate it NT-Petri Nets.

\subsection{Analysis of NT-Petri nets}

Given a transition node, $T$, in a NT-Petri net and a state matrix $s$ which holds the counts of each colored token at each place node, $T$ shall describe the conditions under which it is enabled with a Boolean valued enable function $E_T(s)$.
$E_T(s)$ shall only be dependent on the input place nodes of $T$.
Let $\pi_T(s)$ be the set of possible changes to $s$ performed by $T$.
The set of next possible states is $\{s + \delta | \delta \in \pi_T(s)\}$. In keeping with the definition of Petri nets, no place nodes in a state can have fewer than $0$ tokens.
The state graph of a NT-Petri net start state pair can be found by recursively computing subsequent states allowed by enabled transition nodes, see Algorithm \ref{algo:state_graph}.

\begin{proposition}
If a Petri net's state graph has a cycle, the Petri net must have a cycle. However, if a Petri net has a cycle, the state graph may not have a cycle.
\end{proposition}

This is since if there were no loops, tokens in all places could not be replenished and previous states cannot be achieved from subsequent states. If a Petri net has a loop, a generated state graph from the Petri net may or may not have a loop depending on the start state, since there may exist a loop in the Petri net where tokens can never reach and therefore the corresponding state graph would have no loop.

\begin{algorithm}

    \textbf{Input:} 
    
    \quad ntp : Constant Global NTPetriNet
    
    \quad start\_state: Constant State
    
    \quad state\_graph : Mutable Global StateGraph 
    
    \quad \quad \# Initially only has the  Start\_State as a key
    
    \quad \quad \# and List() as a value
    
    \textbf{Output:}
    
    \quad state\_graph : Mutable Global StateGraph
    
    \textbf{Begin:} 
    
       \For{transition in ntp}{
           \If{transition.enabled(state)}{
                \For{state\_delta in transition.updates(state)}{
                    new\_state = state + state\_delta
                    
                    \If{new\_state not in state\_graph}{
                        state\_graph[state].append\_to\_list(new\_state)
                        
                        state\_graph[new\_state] = List()
                        
                        Compute\_State\_Graph(ntp, new\_state, state\_graph)
                    }
                }
            }
        }
        return state\_graph
        
	\caption{Compute\_State\_Graph}
	\label{algo:state_graph}
\end{algorithm}

The bound for the size of the state graph of a NT-Petri net with up to $t$ tokens, all unique, and $p$ place nodes is 
$$ |State\_Graph| \le (p+1)^t. $$ 

Each token has $p + 1$ choices (any place and the option not to be placed), and since there are $t$ tokens, we get $(p+1)^t$ bound for the number of states. The number of states may be smaller depending on the connectivity and rules of transitions in the NT-Petri net, as a graph.

\section{Concurrent Programs}\label{sec:concurrprog}

Concurrent programs are useful for applications running on hardware capable of parallel execution, applications that use many pieces of hardware, or applications that have very large memory footprints.
Concurrent programs work by having many threads that can be interleaved and executed in any way by an operating system \cite{raynal2012concurrent}. 
A thread is a sequence of instructions to be executed in order. 
A program determines its threads and each thread's sequence of instructions. 
A singe threaded program contains one thread while a concurrent program contains multiple threads. 
Threads depend on resources. A resource is something that the thread uses to accomplish its objective, for example a hard drive, webcam, or CPU time. 
If multiple threads do not share resources correctly, then an unrecoverable state can be reached.
For example, if one thread frees a region of memory while another thread is using that region of memory, the program will be killed by the operating system.

Petri nets are most useful in building and modeling computer programs, not as languages in themselves, but as an API in a standard language.
The Petri net API should have non-blocking subroutines passed into each transition node and executed according to the state of the Petri net and arbitrary constraints may be placed onto the state graph prior to execution to determine if the state graph of the constructed Petri net program is valid as defined by the use case of the application for which the Petri net program is built.

\section{Concurrent Execution of Petri nets}\label{sec:concurrexec}

Petri nets are useful for writing concurrent programs.
We let each token represent a resource and transitions represent functions on those resources. 
If two transition nodes of the Petri net do not share any resource dependency and are both enabled, then they can be safely executed concurrently (in any order or at the same time). We will partition the transition nodes, denote $N_t$, where every pair of partitions do not share incoming place nodes.
Then each partition gets its own thread called a work cluster.
If this is not done, a race condition could occur.
For example, if two work clusters are blocked concurrently on intersecting sets of places, once tokens are deposited in these place nodes, and more than one transition node is enabled in more than one work cluster, then a race between threads occurs and an invalid state that is not on the state graph could be reached.

\begin{figure}
    \centering
    \includegraphics[scale=0.6]{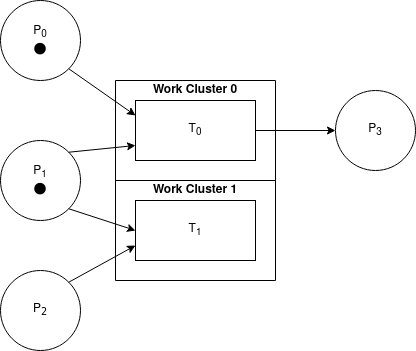}
    \caption{A Petri net with invalid work clusters.}
    \label{fig:pnet_invalid_work}
\end{figure}

\begin{figure}[h]
    \centering
    \includegraphics[scale=0.6]{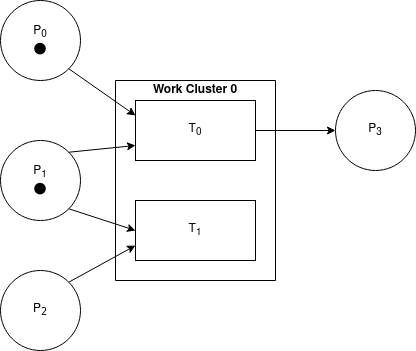}
    \caption{The Petri net of Figure \ref{fig:pnet_invalid_work} with valid work clusters.}
    \label{fig:pnet_valid_work}
\end{figure}

Consider an invalid NT-Petri net example, see Figure \ref{fig:pnet_invalid_work}.
If both $Work Cluster 0$ and $Work Cluster 1$ are blocked on $P_1$, then the receiver of the token at $P_1$ is nondeterministic. If the token goes to $T_0$, then the Petri net can continue its execution to the next state, but if the receiver of the data is $T_1$, then the Petri net will deadlock prematurely as $T_1$ cannot execute due to no token in $P_2$, and $T_0$ cannot execute as there is no token in $P_1$.
Since the transitions $T_1$ and $T_1$ have an intersection of input places, they should be placed in the same work cluster, see Figure \ref{fig:pnet_valid_work}.

Disjoint work clusters may be unioned and the execution of the NT-Petri net will still be valid.
In practice, some Petri nets may execute faster if partitioned into a small number of work clusters due to thread handling overhead present in all operating systems.
Maximal concurrency of a NT-Petri net program may be computed by partitioning the transition nodes into the largest number of valid work clusters. 
When finite, there exists a unique maximal work cluster configuration because transition nodes cannot be shifted between work clusters with the work clusters still being valid. 

This is a type of graph partitioning problem. We want to optimize over all graph partitions minimizing an objective that we set to infinity at invalid partitions. Consider partitions, $P$, of $N_t$, optimize  
$$ \arg\min_{P \ : \ \cup_i P_i = N_t,\ P_i \in P} - \ count( P ) + invalid(P) \times \infty. $$ 
There are many methods for solving a discrete optimization problem such as this as well as many software packages and libraries \cite{steiglitz1998combinatorial}. 
This leads to the conclusion that a NT-Petri net of highly connected place nodes will require a more synchronous execution than a NT-Petri net with transitions that have fewer input intersections.

\section{NT-Petri Net Software Framework}\label{sec:software}

We developed software to implement the above ideas. The software framework is open source and available on 
\href{https://github.com/MarshallRawson/nt-petri-net}{Github.}\footnote{https://github.com/MarshallRawson/nt-petri-net}
This software framework aims to facilitate the construction and execution of software systems modelled as colored NT-Petri nets. 
The framework supports transition nodes, place nodes, and typed tokens.
Each transition node is defined by a set of functions that map defined input firing conditions to possible output tokens. The above theory uses colors but in practice we switch to classes or types in the programmatic sense. A firing condition is defined by requiring a certain number of tokens of a certain type contained at certain input places. A function in a transition node takes a certain number of tokens of given types and outputs certain tokens into output place nodes. Tokens are implemented as unique pointers to dynamically typed objects located on the heap, so moving tokens is efficient and safe. Places are implemented as a set of FIFO queues, with one for each possible type of token it can be given. 

Transitions are implemented as \emph{structs} \cite{klabnik2019rust} with several registered member functions. Each registered member function is assigned one or more input token sets and one or more output token sets. Input and output token sets are implemented as \emph{structs}. Up to one of the registered member functions will be called at a time depending on the input tokens available.

All the parsing of sets of tokens is done in the library itself. It is a purposeful design decision to not expose the list of tokens to the developer in order to minimize parsing errors and maximally leverage the compile-time checks on the developer's code.
The front facing transition API is designed to maximize the compile-time guarantees. 
The resulting NT-Petri net can be executed by as few as one thread or as many threads as there are valid work clusters.

\begin{figure}
\begin{lstlisting}[language=rust, basicstyle=\footnotesize]
mod score {
    ...
    #[derive(ntpnet_macro::TransitionInputTokens)]
    pub struct A { pub a: Vec<String>, }
    #[derive(ntpnet_macro::TransitionInputTokens)]
    pub struct B { pub b: ndarray::Array1<i32>, }   
    #[derive(ntpnet_macro::TransitionInputTokens)]
    pub struct AB {
        pub a: Vec<String>,
        pub b: ndarray::Array1<i32>,
    }
    #[derive(ntpnet_macro::TransitionOutputTokens)]
    pub struct C { pub c: f64, }
    #[derive(ntpnet_macro::TransitionOutputTokens)]
    pub struct D { pub d: f64, }
    #[derive(ntpnet_macro::Transition)]
    #[ntpnet_transition(f: Input(A, B) -> Output(C, D))]
    #[ntpnet_transition(f2: Input2(AB) -> Output2(D))]
    pub struct Score { ... }
    impl Score {
        fn f(&mut self, i: Input) -> Output {
            match i {
                Input::A(A { a } ) => { ... },
                Input::B(B { b } ) => { ... },
            };
        }
        fn f2(&mut self, i: Input2) -> Output2 {
             let AB { a, b } = match i {
                Input2::AB(AB) => AB
            };
            ...
        }
        ...
    }
}
...
fn main() {
    let n = Net::make()
        .place_to_transition("X", "a", "score")
        .place_to_transition("Y", "b", "score")
        .transition_to_place("score", "c", "Z")
        .transition_to_place("score", "d", "Q")
        .add_transition("score", score::Score::maker());
    Reactor::make(n).run();
}

\end{lstlisting}
    \caption{Example NT-Petri net implementation using our software framework written in Rust.}
    \label{fig:code_example}
\end{figure}

\begin{figure}
    \centering
    \includegraphics[scale=0.7]{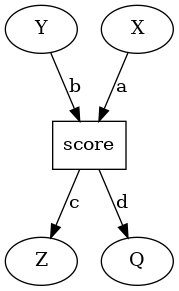}
    \caption{A visualization of the NT-Petri net built by \ref{fig:code_example}.}
    \label{fig:code_example_render}
\end{figure}

Consider an example of a transition with two input places, $X$ and $Y$, and two output places, $Z$ and $Q$. This transition will try to generate a score from the input tokens, but the transition can fire when there is only one token in either $X$ or $Y$, resulting in a token deposited to either $Z$ or $Q$ (depending on internal mechanisms), or if there are tokens in $X$ and $Y$, calculate a score with a different function and place to result into $Q$. Since which score gets calculated in determined by the arrival times of the tokens in $X$ and $Y$, the transitions change to the state graph is nondeterministic. This NT-Petri net is described by the source code in \ref{fig:code_example} and visualized in \ref{fig:code_example_render}.

\section{Smart Pan Tilt Zoom Camera Use Case}\label{sec:cameraex}

\begin{figure}[h]
    \centering
    \includegraphics[scale=0.5]{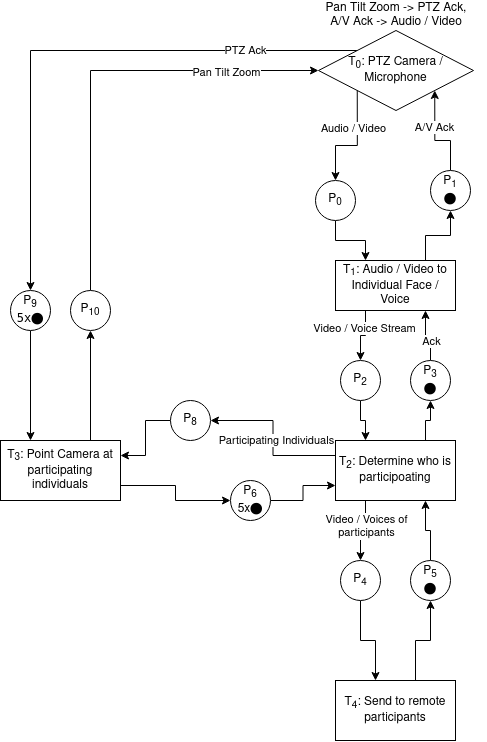}
    \caption{NT-Petri net diagram of a Smart Pan Tilt Zoom Camera System. Squares and diamonds are transition nodes. Circles are place nodes. Black dots are tokens containing data.}
    \label{fig:ptz_example}
\end{figure}

Consider a realistic pan tilt zoom camera and microphone device to allow remote participants to participate in group discussions by providing audio and video of only those speaking with the remote participants.
This is an example of a processing pipeline with feedback. The feedback in this processing pipeline is the pan, tilt, and zoom requests to point the camera at the current speaker.
Typically, a processing pipeline like this with feedback is very difficult to model and show concurrency safety. However, a NT-Petri net provides a concise model of the desired concurrent solution. We visualize and then can easily reason about the NT-Petri net, see Figure \ref{fig:ptz_example}.

We will describe the transitions of this NT-Petri net.
For each transition $T_1$ through $T_5$, when all input places have tokens, the transition is enabled. When firing, a token is removed from each input place and a token is put in each output place. The $T_0$ transition has a special rule determining its firing conditions and resulting set of tokens from each condition. 
This means that the transition can fire if there is a token on $P_{10}$ or $P_1$.
In the case of the former firing condition, a token will be taken $P_{10}$ and put on $P_9$.
In the case of the latter firing condition, a token will be taken $P_1$ and put on $P_0$.
In this solution, $T_0$, $T_1$, $T_2$, $T_4$ are run concurrently, but only advance at the pace of the slowest processing step. $T_3$ has a looser relationship with the processing pipeline, where up to five pan tilt zoom requests (tokens) may be made and not yet fulfilled before execution of the pipeline will possibly wait. While this program can still temporarily block, it also has a finite state graph and therefore can be proven to have correct behavior and no deadlocks.
This shows us that the number of start tokens in certain places can describe the synchronization requirements between certain computations in a concurrent program.

\section{Conclusion}

Modern electronics and computer systems rely on nondeterministic concurrent programs to multitask, for example running a screen and internet connection simultaneously. 
However, concurrent models typically allow for a large number of possible states of the system, most of which are erroneous. These erroneous states are what cause failures and crashes that frustrate technology users. 
Using our framework, we model programs and systems as a Petri nets. 
The state graph of the system can then be computed and checked for invalid states before deploying the system.
A generalization of Petri nets with nondeterministic transition nodes makes the programs more concise, readable, and able to match hardware interfaces whilst still maintaining the ability to optimize the number of threads and solve for the state graph. 
A NT-Petri net program can be optimized by solving for the maximal number of useful threads in its execution since resource requirements of each computation are explicitly stated before execution by the structure of the Petri net.
We describe and give a coded example of our open source software framework that implements these ideas. 
This type of nondeterministic concurrency modelling could be a useful tool for system designers everywhere. 

\section{Future Work}
The two main areas with the largest potential for immediate benefit for systems built as NT-Petri nets are timing introspection and state graph provability.

A system built as a NT-Petri net can record and plot timing of each transition computation vs time wasted in the middle-ware or operating system. The system designer can use this information to make informed design decisions to change how the transitions are connected or change the arrangement of work clusters to increase throughput and decrease latency.

A system built as a NT-Petri net has a state graph that can be computed and verified before execution of the program. However, with large projects, the state graph can become too large. To remedy this problem, the NT-Petri net can be decomposed into components. Then a component's state graphs can be computed and linked together to greatly reduce the size of the state graph as described in \cite{christensen}.

Once a system has both representative timing records and a state graph, timing analysis can be conducted without needing to run the program at all. This eliminates the typical resource and time bottleneck of running and testing systems in the real world. This is extremely useful to system designers attempting to optimize their systems via node rearrangements or collect evidence for timing goals and requirements. 

\section*{Acknowledgements}
The authors acknowledge and thank the reviewers and organizing committee members for their guidance and rich knowledge that helped to substantially improve this paper in many aspects. 

\bibliography{ref}

\end{document}